\documentclass[prl, twocolumn,showpacs,preprintnumbers,amsmath,amssymb]{revtex4}
\usepackage{epsfig}
\usepackage{graphicx}
\usepackage{dcolumn}
\usepackage{bm}
\renewcommand{\vec}[1]{\boldsymbol{#1}}

\def\halft{{\textstyle\frac{1}{2}}}
\def\thirdt{{\textstyle\frac{1}{3}}}

\begin{document}

\title{Solitary and compact-like shear waves in the bulk of solids}

\author{Michel Destrade}
\email{destrade@lmm.jussieu.fr}
   \affiliation{Laboratoire de Mod\'elisation en M\'ecanique
UMR 7607, CNRS, Universit\'e Pierre et Marie Curie,4 place
Jussieu, case 162, 75252 Paris Cedex 05, France}%

\author{Giuseppe Saccomandi}
 \email{giuseppe.saccomandi@unile.it}
\affiliation{Sezione di Ingegneria Industriale, Dipartimento di
Ingegneria dell'Innovazione, Universit\`a degli Studi di Lecce,
73100 Lecce, Italy.}

\date{\today}

\begin{abstract}
We show that a model proposed by Rubin, Rosenau, and Gottlieb [J.
Appl. Phys. \textbf{77} (1995) 4054], for dispersion caused by an
inherent material characteristic length, belongs to the class of
simple materials. Therefore, it is possible to generalize the idea
of Rubin, Rosenau, and Gottlieb to include a wide range of
material models, from nonlinear elasticity to turbulence. Using
this insight, we are able to fine-tune nonlinear and dispersive
effects in the theory of nonlinear elasticity in order to generate
pulse solitary waves and also bulk travelling waves with compact
support.

\end{abstract}


\maketitle

The emergence of significant solitary waves from the equations
governing the motion of solids requires usually a fine balance between
\textit{nonlinearity} and \textit{dispersion}. The first
ingredient finds its natural source in the theory of finite, fully
nonlinear, elasticity, but the source for the second is more
elusive. Often, specific boundary conditions (such as: a thin film
coated on a substrate, the finite interface between phases in an
elastic composite material, the  lateral dimensions of a rod with
free surfaces, etc.) are used because they introduce a
characteristic length, to be compared with the wavelength. For
homogeneous infinite (bulk) solids, the presence of a
microstructure is invoked, recording the presence of inherent
material characteristic lengths due to, for instance: the atomic
lattice cells in a metal, the grain size in an alloy, the
persistence and contour length in a polymer, etc. (it is worth
noting that a similar situation also occurs in fluid dynamics,
when inherent characteristic lengths may be not ignored in models
of turbulence.) Although the standard constitutive equations of
continuum mechanics can accommodate large strains and nonlinear
material responses, they do not in general include the effects of
characteristic material lengths \cite{Maugin}. The introduction of
such information in models of material behavior is a subtle and
difficult problem \cite{KKBT}, and the modifications of the basic
constitutive equations proposed in the literature are often
unsatisfactory because they lead to physically unrealistic and
mathematically ill-posed models \cite{Maugin, RRG}. Finally,
finding a simple connection between nonlinear and dispersive
effects in material modeling is an interesting problem \emph{per
se}, because the balance between these two effects is at the
origin of many fascinating phenomena, and in particular the
propagation of solitons and compactons \cite{RH}.

In 1987 Rosenau \cite{Ros} proposed a regularized functional
expansion of the equations for the dynamics of dense lattices in order
to yield the leading effect of dispersion due to a characteristic
length, via a physically reasonable and mathematically tractable
differential equation.
One of the advantages of this equation over
other micro-mechanical or nonlocal approaches is that it
models dispersion without increasing the number of initial or
boundary conditions of the usual wave equation.
On the same pathway, a three-dimensional phenomenological continuum
model of the Rosenau approach was proposed by Rubin, Rosenau, and
Gottlieb (RRG) in 1995 \cite{RRG}.
One aim of this Letter is to put in perspective that latter model
by showing that it is a special case of the theory of simple
materials in the sense of Noll \cite{TN}, hence generalizing the
original idea of Rosenau to a large class of
material behavior, from nonlinear elasticity to fluid turbulence.
Another objective is to find a clear and direct way to balance
nonlinearity and dispersion in order to generate solitary waves with
compact support in solids.
This balance is attained by relying solely on constitutive arguments,
without any resort to reductive asymptotic expansions;
it is then compared (favorably, as it turns out)
with the results obtained by such methods.

Let the motion of a body be described by a relation
$\vec{x} = \vec{x} (\vec{X}, t)$, where $\vec{x}$ denotes the
current coordinates of a point occupied by the particle of
coordinates $\vec{X}$ in the reference configuration at time $t$.
The deformation gradient $\vec{F} (\vec{X}, t)$
and the spatial velocity gradient $\vec{L} (\vec{X}, t)$
associated with the motion are defined by
$\vec{F} = \text{Grad } \vec{X}$ and
$\vec{L} = \text{grad } \vec{v}$
($\vec{v} = \partial \vec{x} / \partial t$), respectively
(in this Letter, grad and div denote respectively the gradient and
divergence operators with respect to the current position $\vec{x}$,
and Grad and Div denote the corresponding operators with respect to
$\vec{X}$.)
The mathematical theory of simple materials is well-grounded from the
mathematical and thermomechanical points of view \cite{TN};
there, the Cauchy stress $\vec{T}$ is determined by the
whole history of the deformation gradient,
\begin{equation}
\vec{T} = \mathcal{G}_{s=0}^\infty \left(\vec{F}(t-s) \right),
  \label{simple}
\end{equation}
where $\mathcal{G}$ is the constitutive functional. We now
summarize the RRG model, which itself follows the axiomatic
approach to continuum mechanics introduced by Green and Naghdi
\cite{Green}. Let us introduce the specific entropy  per unit mass
$\eta$, the specific rate of internal production of entropy $\xi$,
the entropy flux per unit of present area $\vec{p}$, and the
specific Helmholtz free energy $\psi$. The set $\{\eta, \xi,
\vec{p}, \psi, \vec{T} \}$ must satisfy, for all possible
thermomechanical process, the restrictions coming from the balance
of angular momentum, the energy equation, and from a suitable
statement of the second law of thermodynamics. The central idea of
Rubin et al. \cite{RRG} is to add terms $\{\psi_2, \vec{T}_2 \}$,
\textit{modeling material dispersion}, to the Cauchy stress tensor
and the free energy: $\vec{T} = \vec{T}_1 + \vec{T}_2$, $\psi =
\psi_1 + \psi_2$, say, while the other quantities $\{\eta, \xi,
\vec{p} \}$ and the associated constitutive equations stay
unchanged. They show that the balance of angular momentum and the
energy equation then lead to $\vec{T}_2  =  \vec{T}_2^T$ and
\begin{equation}
  \rho \dot{\psi}_{2} - \vec{T}_2 \cdot \vec{D} = 0,
  \label{dis4}
\end{equation}
respectively, where $\rho$ is the mass density per unit volume,
a superposed dot denotes the material time derivative
(at $\vec{X}$ fixed), and $\vec{D} = (\vec{L} + \vec{L}^T)/2$
is the stretching tensor.
Further, the RRG model considers that $\psi_2$ is a single-valued
function: $\psi_2=\psi_2(\delta)$, where
$\delta \equiv \vec{D\cdot D}$ is an isotropic invariant;
then from \eqref{dis4} follows that
\begin{equation}
\vec{T}_2 =
 \rho \psi'_2(\delta)
  \left[ \text{grad } \dot{\vec{v}} + (\text{grad } \dot{\vec{v}})^T
             + 2 \vec{L}^T \vec{L} - 4\vec{D}^2 \right].
  \label{D}
\end{equation}

Our first result is to show that $\vec{T}_2$ is a special case
of the Cauchy stress $\vec{\sigma}$ associated with those
non-Newtonian simple fluids denoted in the literature as
 \textit{second-grade fluids} \cite{TN, Raja},
\begin{equation}
 \vec{\sigma} =
  \nu \vec{A}_1 + \alpha_1\vec{A}_2 + \alpha_2\vec{A}_1^2,
 \label{TD}
\end{equation}
where $\nu$ is the classical Newtonian viscosity,
$\alpha_1$, $\alpha_2$ are the microstructural coefficients
(which may be taken as constants or as functions,
for example of $\delta$),
and the first two Rivlin-Ericksen tensors $\vec{A}_1$, $\vec{A}_2$ are
defined by:
$\vec{A}_1 = 2 \vec{D}$,
$\vec{A}_2 =\dot{\vec{A}}_1 + \vec{A}_1 \vec{L} + \vec{L}^T \vec{A}_1$.
Indeed, recall that
$\vec{L} = (\text{Grad }\vec{v}) \vec{F}^{-1}
                  = \dot{\vec{F}} \vec{F}^{-1}$,
from which follows that
\begin{equation}
\dot{\vec{L}} =
   (\text{Grad } \dot{\vec{v}}) \vec{F}^{-1}
     + \vec{L} \vec{F} \dot{(\vec{F}^{-1})}
 =  \text{grad } \dot{\vec{v}} - \vec{L}^2,
\end{equation}
(where the last equality comes from the material time derivative
of the identity $\vec{F} \vec{F}^{-1} = \vec{1}$). Substituting
into $\dot{\vec{A}}_1 = \dot{\vec{L}} + \dot{\vec{L}}^T$ and into
the definition of $\vec{A}_2$, we find that the tensors \eqref{D}
and \eqref{TD} coincide when $\nu =0$ and $\alpha_1(\geq0) = -
\alpha_2 = \rho \psi'_2$. We conclude that when $\vec{T}_1$ is in
the material class \eqref{simple} then the total Cauchy stress
tensor $\vec{T} = \vec{T}_1 + \vec{T}_2$ of the corresponding RRG
model is also that of a simple material. This observation has many
interesting consequences. For example, if $p$ is the Lagrange
multiplier associated with the constraint of incompressibility
($\text{det }\vec{F} = 1, \text{tr } \vec{D} = 0$), then the
classical choice of the Navier-Stokes stress tensor $\vec{T}_1 =
-p \vec{1}+\nu \vec{D}$ for the RRG model leads to those
incompressible homogeneous fluids of second grade for which
$\alpha_1 + \alpha_2 = 0$. We note that the equations of motion
associated with those fluids, in the general and some special
case, have been found \cite{Busu99}
to coincide with the equations derived by
other means (asymptotic expansions, Lagrangian means, average
Euler equation, etc.) in several models of turbulence and shallow
water theory \cite{turb}. It is now clear that the presence of
the RRG characteristic length is the reason of such connections.

Let us consider the case of the RRG model when $\vec{T}_1$
corresponds to a hyperelastic, incompressible, isotropic
\textit{solid} \cite{TN},
\begin{equation}
 \vec{T}_1 =
     - p \vec{1}
      + 2(\partial \Sigma / \partial I_1) \vec{B}
       - 2 (\partial \Sigma / \partial I_2) \vec{B}^{-1}.
 \label{TE}
\end{equation}
Here $\vec{B} = \vec{F} \vec{F}^T$, and
$\Sigma = \Sigma(I_1, I_2)$ is the strain energy density, a function
of the invariants $I_1 = \text{tr }\vec{B}$ and
$I_2 = \text{tr }\vec{B}^{-1}$.
Then the RRG model is a special case of the so-called
\textit{solid of second grade} \cite{Fos, DS}.
A rectilinear shearing motion is defined by
\begin{equation}
\vec{x}= \left[ X + u(Y,t) \right] \vec{e}_1
                   + Y \vec{e}_2 + Z\vec{e}_3,
  \label{d1}
\end{equation}
where ($\vec{e}_1, \vec{e}_2, \vec{e}_3$) is a fixed orthonormal triad.
Then
$\vec{B} = \vec{1}
              + u_Y^2 \vec{e}_1 \otimes \vec{e}_1
                + u_Y (\vec{e}_1 \otimes \vec{e}_2
                  + \vec{e}_2 \otimes \vec{e}_1)$,
$\vec{B}^{-1} = \vec{1}
                  + u_Y^2 \vec{e}_2 \otimes \vec{e}_2
                    - u_Y (\vec{e}_1 \otimes \vec{e}_2
                      + \vec{e}_2 \otimes \vec{e}_1)$,
$\vec{A}_1 =
  u_{Yt} (\vec{e}_1 \otimes \vec{e}_2 + \vec{e}_2 \otimes \vec{e}_1)$,
and
$\vec{A}_2 =
  u_{Ytt}(\vec{e}_1 \otimes \vec{e}_2 + \vec{e}_{2} \otimes \vec{e}_1)
    + 2 u_{Yt}^2 \vec{e}_2 \otimes \vec{e}_2$.
Also, $I_1 = I_2 = 3 + u_Y^2$ and $\delta =u_{Yt}^2$.
The equations of motion in the absence of body
forces ($\text{div } \vec{T} = \rho \dot{\vec{v}}$) reduce to the
three scalar equations \cite{DS}:
\begin{equation}
-\frac{\partial p}{\partial X}
  + \frac{\partial T_{12}}{\partial Y}
   = \rho  \frac{\partial^2 u}{\partial t^2},
 \quad
\frac{\partial T_{22}}{\partial Y} = 0,
 \quad
- \frac{\partial p}{\partial Z} = 0,
 \label{Eq}
\end{equation}
for which we used \eqref{D}, or its equivalent form:
$\vec{T}_2 = \rho \psi'_2 (\vec{A}_2 - \vec{A}_1^2)$.
The third equation gives $p=p(X,Y,t)$ and then the
cross-differentiation of the first and second equations leads to
$p(X,Y,t) =
  - 2 \Sigma_2 u_Y^2
     + \rho \psi_2' u^2_{Yt}+ p_1(t) X + p_0(t)$,
where $p_0$, $p_1$ are arbitrary functions of $t$ alone and
$\Sigma_i$ is the partial derivative of $\Sigma$ with respect to
$I_i$.
We are left with a single determining equation for
the displacement $u(Y,t)$:
$p_1 + [2(\Sigma_1 + \Sigma_2) u_{Y} + \rho \psi_2' u_{Ytt}]_Y
   = \rho u_{tt}$ or, taking the derivative with
respect to $Y$, a single equation for the \textit{strain}
$U \equiv u_{Y}$,
\begin{equation}
 \left[ \hat{\mu} U + \hat{\beta} U_{tt} \right]_{YY} = \rho U_{tt},
\, \hat{\mu} = 2(\Sigma_1 + \Sigma_2), \,
 \hat{\beta} = \rho \psi_2'.
  \label{eqgen}
\end{equation}
Here $\hat{\mu} = \hat{\mu}(U^2) >0$ is the generalized shear modulus
of nonlinear elasticity \cite{Carr, Carr2}; it is a constant
(Lam\'e's second constant) in linear elasticity, and also in
nonlinear elasticity for the special cases of the neo-Hookean
(``geometric'' nonlinearity) and of the Mooney-Rivlin
(``geometric'' and ``physical'' nonlinearities) forms of the
strain energy $W$ ($= C_{10} (I_1 - 3)$, $ C_{10}(I_1-3) + C_{01}
(I_2-3)$, respectively). However in general, a nonlinear
constitutive equation for the solid yields a non-constant
$\hat{\mu}$; similarly, a general dispersive material behavior is
attested by a non-constant function $\hat{\beta} =
\hat{\beta}(U_t^2) > 0$ (the case where $\hat{\mu}$ and
$\hat{\beta}$ are both constants is considered elsewhere
\cite{DS}). To illustrate some features of the RRG model, we make
the following constitutive choices: $\hat{\mu} (U^2) = \mu_0 +
\mu_1 U^2$ and $\hat{\alpha} (U_t^2) = \rho(\beta_0 + \beta_1
U_t^2)$, where $\mu_0 >0$ is the infinitesimal shear modulus of
the solid. We seek a travelling wave solution: $U = V(\xi)$, $\xi
= Y - c t$, with drop boundary conditions: $V^{(n)} (\pm \infty) =
0$, $n \ge 0$. Then successive integrations of the reduced
governing equation lead to
\begin{equation}
\rho c^4 \beta_1 (V')^4 + 2 \rho c^2 \beta_0 (V')^2
 + [\mu_1 V^2 - 2(\rho c^2 - \mu_0)]V^2 = 0.
 \label{reduced}
\end{equation}

For a linear variation of the dispersion free energy $\psi_2$
with the isotropic invariant $\delta = \vec{D} \cdot \vec{D}$,
we have $\beta_1 = 0$ and the following \textit{solitary wave}
solution:
\begin{equation}
 V(\xi) =
   \sqrt{2 \dfrac{\rho c^2 - \mu_0}{\mu_1}}
     \text{sech } \sqrt{\dfrac{\rho c^2 - \mu_0}{\rho c^2}}
                   \dfrac{\xi - \xi_0}{\sqrt{\beta_0}},
  \label{power4}
\end{equation}
where $\xi_0$ is arbitrary.
This wave is supersonic with respect to the speed of
an infinitesimal shear wave but $c$ is otherwise arbitrary.
Figure 1 shows its profile for several values of $\rho c^2 / \mu_0$.
\begin{figure}
\includegraphics[width=50mm, height=50mm]{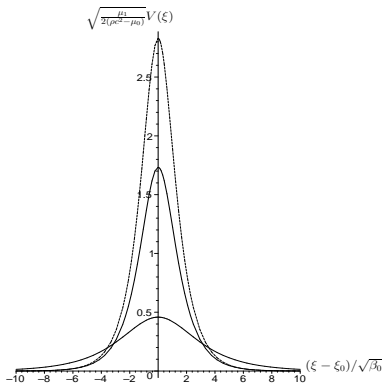}
\caption{\label{fig1} Profile of the solitary wave when $\rho
c^2/\mu_0 = 1.1$ (thick curve), 2 (thin curve), 3 (dashed curve).}
\end{figure}

Now, we know from previous results \cite{SAC} that \eqref{reduced}
has no compact-like solutions when $\beta_0 \beta_1 \ne 0$.
On the other hand, when $\beta_0 = 0$, $\beta_1 \ne 0$, it is of
the form
\begin{equation}
 (V')^4 = (\gamma_0 - \gamma_1 V^2) V^2,
   \label{power9}
\end{equation}
(where $\gamma_0 = 2(\rho c^2 - \mu_0)/(\rho c^4 \beta_1)$
and $\gamma_1 = \mu_1/(\rho c^4 \beta_1)$),
the right hand-side of which has a double zero (at $V=0$) and
two simple zeros (at $V = \pm \sqrt{\gamma_0 / \gamma_1}$).
This observation alone is sufficient to conclude that \eqref{power9}
admits a \textit{compact wave} solution.
As it happens, \eqref{power9} can be integrated in terms of special
functions.
Indeed, introducing the hypergeometric function
$\text{Hyp}_2\text{F}_1 [a, b, c, z]$,
solution to the second-order linear differential equation
$z(1-z)y'' + [c - (a + b + 1)z]y' - a b y = 0$, we find the
following identity,
\begin{equation}
 \int (u^2 - u^4)^{-1/4} \text{d}u =
   2 (u^2)^{1/4}
    \text{Hyp}_2\text{F}_1
       \left[\frac{1}{4}, \frac{1}{4},
          \frac{5}{4}, u^2 \right].
 \label{power12}
\end{equation}
We call $\mathcal{I}(u)$ that function and observe that it is
defined on the interval $0 \le u \le \pi/\sqrt{2}$,
where it increases in a monotone way from 0 to 1.
Then we manipulate \eqref{power9} to give
\begin{equation}
 \sqrt{\gamma_1 / \gamma_0} V(\xi) =
   \mathcal{I}^{-1} [\gamma_1^{1/4} (\xi - \xi_0)],
     \label{power13}
\end{equation}
where $\xi_0$ is arbitrary. Hence we can build a \textit{weak}
solution with \textit{finite} support measure $\sqrt{2} \pi
\gamma_1^{-1/4}$, defined by:
\begin{align}
\sqrt{\gamma_1 / \gamma_0} V(\xi) & =
   \mathcal{I}^{-1} (\gamma_1^{1/4} \xi)
&& \text{on } [0, \pi/a],
\notag \\
 & = \mathcal{I}^{-1} [\gamma_1^{1/4} (\sqrt{2}\pi -\xi)]
&& \text{on } [\pi/a, \sqrt{2}\pi / a],
\end{align}
and zero \textit{everywhere else}, see Figure 2. (Here $a=(4
\gamma_1)^{1/4}$). This wave shares some characteristics with the
solitary wave \eqref{power4}: it is supersonic, its amplitude is
proportional to $\sqrt{\gamma_0 / \gamma_1} = \sqrt{2(\rho c^2 -
\mu_0)/\mu_1}$, growing with the speed; in contrast, its width
evolves as $\gamma_1^{-1/4}$, which is proportional to $c$. We
emphasize that this compact wave is derived solely on the basis of
constitutive arguments, by taking the generalized shear modulus
$\hat{\mu}$ linear in the squared shear (nonlinear effect) and the
free energy density $\psi_2$ proportional to $\delta^2$
(dispersive effect). Previous attempts at finding compact waves
relied on ad hoc asymptotic reductive approaches \cite{FLS} and to
the best of our knowledge, a bulk compact wave has never been
generated in the framework of nonlinear elasticity \cite[p.
91]{Maugin}
\begin{figure}
\includegraphics[width=50mm, height=50mm]{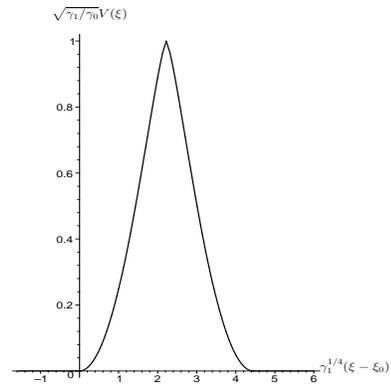}
\caption{\label{fig} Profile of the compact wave.}
\end{figure}

Finally we look at reductive asymptotic expansions of the
strain equation of motion \eqref{eqgen}.
We consider that the generalized shear modulus is slightly
nonlinear, $\hat{\mu} = \mu_0(1 + \epsilon U^2)$ where $\epsilon$ is
a small parameter, and that the dispersion is also of order
$\epsilon$: for instance, when $\psi_2$ is linear in $\delta$,
we take: $\hat{\beta} = 2 (\rho a^2) \epsilon $, where $a$ is a
constant with the dimensions of a length.
With the standard semi-characteristic variable stretching
transformation  \cite{JK}: $U(Y,t) = v(\eta, \epsilon \tau)$, where
 $\tau = [\mu_0 / (\rho a^2)]^{1/2} t$ and
$\eta = Y/a - \tau$ are dimensionless variables, we find that at
order $\epsilon$, \eqref{eqgen} is the modified KdV equation,
\begin{equation}
 v_\tau + \halft (v^3)_\eta + v_{\eta \eta \eta} = 0,
 \label{evo}
\end{equation}
for which the solitonic solution is a \textit{sech}-wave as in
(\ref{power4}). When $\psi_2$ is proportional to $\delta^2$, we
take $\hat{\beta}$ in the form $\hat{\beta} = 2 (\rho^2 a^4/\mu_0)
\epsilon U_t^2$, and find
\begin{equation}
 v_\tau +  \halft (v^3)_\eta +  \thirdt (v_\eta^3)_{\eta \eta} = 0,
 \label{evo2}
\end{equation}
an equation similar to the $K(3,3)$ KdV equation \cite{RH,
Rose96}. Although \eqref{evo} and \eqref{evo2} might have rich
implications, we do not study them further because the generality
of  \eqref{eqgen} has clearly been lost in the process of deriving
them. In particular, the traveling equation \eqref{reduced} is
just as rewarding, and is obtained exactly, without the need for a
time-stretching variable, a moving frame, and an asymptotic
expansion where higher-order terms are simply ignored.

In conclusion, we believe that this investigation demonstrates the
usefulness, the versatility, and ultimately the beauty of the RRG
model. Although the RRG model originates from a microstructural
model \cite{Ros}, it can now be apprehended as a purely
phenomenological approach to dispersion, applicable to the whole
class of simple materials. It overcomes the usual problems of
nonlocal theories, it unifies several results in the literature,
and it provides a natural and detailed study of localization of
traveling waves in elastic materials. Yet several topics raised
here merit further scrutiny: is the solitary wave exhibited a
\textit{soliton}? is the compact wave a \textit{compacton}? Also,
the analysis relied on \textit{specific} constitutive equations.
Hence for the nonlinear elasticity, the generalized shear modulus
was that of a solid having strain energy density in the form
\cite{Carr2} $\Sigma = C_{10}J_1 + C_{01}J_2 + C_{20} J_1^2 +
C_{11}J_1 J_2 + C_{02}J_2^2$ where $J_1 = I_1 - 3$ and $J_2 =
I_2-3$ (so that $\mu_0 = 2(C_{10}+C_{01})$ and $\mu_1 =
4(C_{20}+C_{11}+C_{02})$). Although this, and higher-order,
truncation of the polynomial expansion of $\Sigma$ are highly
popular in experimental nonlinear elasticity, in curve fitting,
and in finite element packages, it should be acknowledged that the
determination of the ``best'' $C_{ij}$ is a messy procedure
\cite{OgSS04}. Eventually, it might be of greater (and safer)
interest to consider other types of nonlinear strain energy
densities, such as those arising in biological soft tissues, which
incorporate the effect of strain hardening. Similar remarks hold
for the nonlinear dispersion, for which the free energy was taken
as $\psi_2 = \alpha_0 \delta + (\alpha_1/2)\delta^2$.

\bigskip

This work was supported by a S\'ejour Scientifique de Haut Niveau
from the French Minist\`ere des Affaires Etrang\`eres, by GNFM and
by MIUR PRIN 2004.

%%%%%%%%%%%%%%%%%%%%%%%%%%%%%%

%%%%%%%%%%%%%%%%%%%%%%%%%%%%%%%%%%%%%%%%%%%%%%%%%%%%%

\end{document}